\documentclass[apj]{emulateapj}
\usepackage{graphicx}

\shorttitle{Survey of H$_3^+$ toward the Galactic Center II.}
\shortauthors{Goto et al.}

\begin{document}

\title{Absorption Line Survey of H$_3^+$ toward the Galactic Center
  Sources II. \\ Eight Infrared Sources within 30~pc of the Galactic
  Center\altaffilmark{1}}

\author{Miwa Goto,\altaffilmark{2}
        Tomonori Usuda,\altaffilmark{3}
        Tetsuya Nagata,\altaffilmark{4}
        T. R. Geballe,\altaffilmark{5}
        Benjamin J. McCall,\altaffilmark{6}
        Nick Indriolo,\altaffilmark{6}
        Hiroshi Suto,\altaffilmark{7}
        Thomas Henning,\altaffilmark{2}
        Christopher P. Morong,\altaffilmark{8}
\protect\and Takeshi Oka\altaffilmark{8}}

\email{mgoto@mpia.de, t-oka@uchicago.edu}

\altaffiltext{1}{Based on data collected at Subaru Telescope,
                 operated by the National Astronomical Observatory of Japan}


\altaffiltext{2}{Max-Planck-Institut f\"ur Astronomie, K\"onigstuhl 17,
                 Heidelberg D-69117, Germany}

\altaffiltext{3}{Subaru Telescope, 650 North A`ohoku Place, Hilo,
                 HI 96720}

\altaffiltext{4}{Department of Astronomy, Kyoto University, Kyoto,
                 606-8502, Japan}

\altaffiltext{5}{Gemini Observatory,
                 670 North A`ohoku Place, Hilo, HI 96720.}

\altaffiltext{6}{Department of Astronomy and Department of
                 Chemistry, University of Illinois at Urbana-Champaign,
                 Urbana, IL 61801-3792.}

\altaffiltext{7}{National Astronomical Observatory of Japan, Osawa,
                 Mitaka, Tokyo 181-8588, Japan}

\altaffiltext{8}{Department of Astronomy and Astrophysics,
                 Department of Chemistry, and Enrico Fermi Institute,
                 University of Chicago, Chicago, IL 60637.}

\begin{abstract}

Infrared absorption lines of $\mathrm{H}_{3}^{+}$, including the
metastable $R$(3,3)$^l$ line, have been observed toward eight bright
infrared sources associated with hot and massive stars located in and
between the Galactic Center Cluster and the Quintuplet Cluster 30~pc
to the east. The absorption lines with high velocity dispersion arise
in the Galaxy's Central Molecular Zone (CMZ) as well as in foreground
spiral arms. The temperature and density of the gas in the CMZ, as
determined from the relative strengths of the $\mathrm{H}_{3}^{+}$
lines, are $T$=200~--~300~K and $n\leq$50~--~200~cm$^{-3}$. The
detection of high column densities of $\mathrm{H}_{3}^{+}$ toward all
eight stars implies that this warm and diffuse gaseous environment is
widespread in the CMZ. The products of the ionization rate and path
length for these sight lines are 1000 and 10 times higher than in
dense and diffuse clouds in the Galactic disk, respectively,
indicating that the ionization rate, $\zeta$, is not less than
10$^{-15}$~s$^{-1}$ and that $L$ is at least on the order of 50~pc.
The warm and diffuse gas is an important component of the CMZ, in
addition to the three previously known gaseous environments: (1) cold
molecular clouds observed by radio emission of CO and other molecules,
(2) hot ($T~=$~10$^4$~--~10$^6$~K) and highly ionized diffuse gas
($n_e~=$~10~--~100~cm$^{-3}$) seen in radio recombination lines, far
infrared atomic lines, and radio-wave scattering, and (3) ultra-hot
($T~=$~10$^7$~--~10$^8$~K) X-ray emitting plasma. Its prevalence
significantly changes the understanding of the environment of the
CMZ. The sight line toward GC~IRS~3 is unique in showing an additional
$\mathrm{H}_{3}^{+}$ absorption component, which is interpreted as due
to either a cloud associated with circumnuclear disk or the
``50~km~s$^{-1}$ cloud'' known from radio observations. An infrared
pumping scheme is examined as a mechanism to populate the (3,3)
metastable level in this cloud.

\end{abstract}



\section{Introduction}

\subsection{The Central Molecular Zone}

The region of the Galaxy within 200~pc of the center, known as the
Central Molecular Zone (CMZ), contains 10\% of the Galactic
interstellar molecular mass \citep[and references
therein]{mor96,gen94,mez96}. The interstellar medium (ISM) in the
CMZ is exceptional in many other ways. It is highly turbulent,
pervaded by an intense magnetic field \citep[for a review;
][]{mor06}, contains a multitude of massive stars
\citep[e.g.,][]{mon92,cot98,mun06} and super-massive clusters
\citep[e.g.,][for a short review]{nag90,nag95,fig04}, and also is
home to numerous highly energetic objects ranging from supernova remnants
to cataclysmic variables \citep[CVs, e.g.,][]{mun03}. The particle
densities in its dense molecular clouds are of order
10$^4$~cm$^{-3}$ and higher. The volume filling factor,
\emph{f$_{\rm v}$}, of dense molecular clouds has previously been
estimated to be $\sim$ 10\% \citep{mor96}, but is more likely $\sim$
1\% \citep{oka05} in view of the observed visual extinction of
$\mathit{A_V}$~=~25~--~40~mag \citep{cot00}.

If dense molecular clouds only occupy 1\% of the CMZ, what
interstellar environments make up the remaining 99\%? There have
been some reports of lower density ``diffuse'' molecular gas in
the intercloud region. Detailed analysis of the $\mathit{J}$ = 1
$\rightarrow$ 0 emission lines of CO and C$^{18}$O by
\citet{dah98}, and the $\mathit{J}$~=~2~$\rightarrow$~1 and
$\mathit{J}$ = 1 $\rightarrow$ 0 emission lines of CO by
\citet{oka98a} have demonstrated the presence of gas with
densities of $\sim$ 10$^{2.5}$ cm$^{-3}$. Indeed \citet{oka98a}
conclude that CO emission from the Galactic center (GC) is
dominated by emission from gas of that density.  The volume
filling factor of \emph{f$_{\rm v}$} $\sim$ 0.1 claimed by
\citet{mor96} may include such gas. In this paper we present
evidence for an even lower density molecular environment in
which H$_2$ must still be plentiful but where CO is not
abundant. In hindsight, the radio absorption lines with high
velocity dispersion of OH \citep[e.  g.][]{bol64}, H$_2$CO
\citep[e. g.][]{pal69}, HCO$^+$ \citep{lin81}, and CO
\citep{oka98b} may originate in this newly reported environment.

A second candidate to fill the intercloud medium is the ultra-hot
plasma ($T$ $\sim 10^7$ -- 10$^8$~K) observed by its diffuse X-ray
emission at energies of 0.5 -- 10~keV. The radiation from the plasma 
is observed in all sightlines to the CMZ
\citep[e.g.,][]{koy89,yam93,koy96}. \citet{laz98} suggested that the
highly ionized regions with \emph{n$_e$} $\sim 10$ cm$^{-3}$ and $T$
$\sim 10^6$~K that are required to explain the $\lambda^2$ dependent
radio-wave scattering toward the GC are located at the
photon-dominated interfaces between the molecular clouds and this
ultra-hot X-ray emitting plasma. They proposed that high density
molecular clouds with a volume filling factor of perhaps \emph{f$_{\rm
v}$} $\sim$ 0.1 are surrounded by the scattering electron gas and the
rest of the space is filled with this plasma (see their Fig. 9).

For an environment such as that described above, which is so dominated
by ionized gas, ``Central Plasma Zone" rather than CMZ would be a more
appropriate term for the region. For a variety of reasons the
ultra-hot and hot plasma cannot occupy the same regions where
molecules abound. How those different categories of gas are
distributed and how they coexist in the CMZ is yet to be
understood. Observations of a new astrophysical probe of the CMZ, the
infrared absorption spectrum of H$_3^+$ \citep{geb96}, are beginning
to shed fresh light on these questions \citep{bol06}. Earlier
measurements of this molecular ion in the Galactic center by
\citet{geb99}, \citet{got02}, and \citet{oka05} were made along only a
very few sight lines. In this paper we report and analyze observations
of seven transitions of H$_3^+$ along eight sight lines, which provide
a more comprehensive diagnostic of the extent and nature of the
H$_3^+$ - containing gas in the CMZ.

\begin{table*}[bht]
\begin{center}
\tablewidth{\textwidth}
\scriptsize
  \caption{Target list.}\label{tb1}
  \begin{tabular}{lcccccll}
\hline \hline
Source                &  R.A.      &  Dec.       &   $l$    & $b$      &$L$/$L'$& Other Names               & Reference    \\
                      &     (J2000)&     (J2000) & [\arcdeg]& [\arcdeg]& [mag]  & in SIMBAD                 & 1, 2, 3.     \\
\hline

GC~IRS~21        \dots& 17:45:40.2 & $-$29:00:31 &$-$0.0561 &$-$0.0468 &        &NAME SGR~A~IRS~21, GCIRS~21 & 1, 2, 3.     \\
GC~IRS~3        \dots & 17:45:39.8 & $-$29:00:24 & $-$0.0552&$-$0.0448 &  5.3   &BHA~11, GCIRS~3             & 1, 2, 3.     \\
GC~IRS~1W        \dots& 17:45:40.4 & $-$29:00:27 &$-$0.0549 &$-$0.0471 &  5.5   &NAME SGR~A~IRS~1W, GCIRS~1W & 1, 2, 3.     \\
NHS~21           \dots& 17:42:53.8 & $-$28:51:41 &  +0.0992 &$-$0.0553 &  4.62  &qF~577, [NHS93]~21          & 4, 5, 6.     \\
NHS~22           \dots& 17:46:05.6 & $-$28:51:41 &  +0.1199 &$-$0.0482 &  6.0   & [NHS93]~22                 & 4, 5, 6.     \\
NHS~42           \dots& 17:46:08.3 & $-$28:49:55 &  +0.1481 &$-$0.0426 &  6.05  & qF~578, [NHS93]~42         & 4, 5, 6.     \\
NHS~25           \dots& 17:46:15.3 & $-$28:50:04 &  +0.1592 &$-$0.0657 &  6.2   &NAME PISTOL STAR, [NHS93]~25& 4, 5, 6.     \\
GCS~3-2       \dots   & 17:46:14.7 & $-$28:49:41 &  +0.1636 &$-$0.0606 &  2.5   & GCS~3-2, [NHS93]~24        & 4, 5, 6.     \\
\hline
    \end{tabular}

\tablecomments{Reference: 1. \citet{kra95}, 2. \citet{ott99}, 3.
\citet{tan06}, 4. \citet{nag90}, 5. \citet{nag93}, 6. \citet{fig99}.
}

\end{center}
\end{table*}

\subsection{H$_3^+$ as an Astrophysical Probe}

The crucial information obtained from the observations of
H$_3^+$ reported here is the product of the ionization rate of
molecular hydrogen \emph{$\zeta$}, and the aggregate length of
the line of sight through the H$_3^+$ - containing clouds,
$L$. The simple chemistry of H$_3^+$ allows us to obtain this
product from the observed total H$_3^+$ column densities
\citep{oka05,oka06b}, as described below, under the assumptions
that (1) there is a steady-state balance between creation and
destruction of H$_3^+$ and (2) the dominant supplier of
electrons is carbon, all of which is singly ionized.

Creation of H$_3^+$ begins with ionization of H$_2$.  Because in
interstellar regions where H$_2$ is plentiful the ion-neutral
reaction H$_2$ + H$_2^+$ $\rightarrow$ H$_3^+$ + H is rapid, the
limit to the formation of H$_3^+$ is the much slower ionization
rate of molecular hydrogen,

\[
{\rm H_2} + {\rm R} \rightarrow {\rm H_2^+} + e + {\rm R}^{\prime}.
\]

\noindent The ionizing rays, R, are dominated by cosmic rays in the
Galactic disk \citep[][and references therein]{ind07}, but in the
CMZ the ionization by X-rays and EUV radiation may be significant.
The production rate per unit volume of H$_3^+$ is thus given by
\emph{$\zeta$}$n$(H$_2$), where \emph{$\zeta$} is the effective
ionization rate of H$_2$ including all the aforementioned
contributors. In steady state, production of H$_3^+$ balances its
destruction, which is dominated by dissociative recombination with
electrons,

\[
{\rm H_3^+} + {\rm e} \rightarrow {\rm 3H~~or~~H_2} + {\rm H},
\]

\noindent
and one obtains \citep{geb99}

\[
\zeta n({\rm H_2}) = k_{\rm e}n({\rm H_3^+})n({\rm e}),
\]

\noindent
where $k_e$ is the electron recombination rate constant,
whose temperature dependent value is $\sim$ 10$^{-7}$ cm$^3$
s$^{-1}$ \citep{mcc04}. Replacing the electron density with the
atomic carbon density after depletion gives

\[
{n({\rm e})}/{n({\rm H_2})} = ({2}/{f}) \left[{n_{\rm C}}/{n_{\rm
H}}\right]_{\rm SV} R_X,
\]

\noindent where $R_X$ = 3 -- 10 is the factor increase in relative
carbon abundance at the Galactic center over the solar vicinity,
$[n_{\rm C}/n_{\rm H}]_{\rm SV}$ \citep{sod95}, and $f = 2n({\rm
H_2}) /n_{\rm H}$ is the fractional density of molecular hydrogen
relative to the total number of hydrogen atoms, $n_{\rm H} = 2n({\rm
H_2}) + n({\rm H})$. Using $N$(H$_3^+$)~=~$L$$n$(H$_3^+$) because of
the constancy of $n$(H$_3^+$) \citep[e.g.,][]{oka06b}, we obtain the
equation central to the analysis in this paper relating the product
\emph{$\zeta$L} to the total H$_3^+$ column density and other
measurables,

\begin{equation}
          \zeta L = 2 k_{\rm e} \left[{n_{\rm C}}/{n_{\rm H}}\right]_{\rm
SV} R_X N({\rm H_3^+})_{total}/f.
\end{equation}

As discussed by \citet{oka05}, in the Galactic center the spectrum of
H$_3^+$ provides direct information about the temperature and density
of the gas. There the relative population of H$_3^+$ in the ($J$, $K$)
= (3,3) metastable level, which is 361 K above the lowest (1,1) level
(note that $J$ =~0 does not exist in the ground vibrational state), is
measurable and is a useful thermometer for temperatures of $\sim$ 100
-- 1000~K. Observations of the fractional population in the (2,2)
unstable level, from which H$_3^+$ decays to the (1,1) level by
spontaneous emission in 27 days \citep{pan86,nea96}, provides a good
measure of cloud densities less than several hundred per cm$^{3}$. The
lifetime of H$_3^+$ in diffuse clouds is $1/(k_{\rm e} n_{\rm e})
\approx 10^9$~s, which is two orders of magnitude longer than the
collisional timescale with molecular hydrogen. The rotational
temperature of H$_3^+$ is therefore maintained via collisions with the
ambient H$_2$. The thermalization of H$_3^+$ in interstellar clouds
has been modeled by \citet{oka04}, who provide H$_3^+$ population
ratios $n$(3,3)/$n$(1,1) and $n$(3,3)/$n$(2,2) plotted as functions of
temperature and density (see their Figs.~2-4).

\section{Observations}

Bright infrared sources corresponding to dust shells surrounding
luminous stars or to hot and bright stars with a dearth of
photospheric absorption features were selected from the Sgr~A$^\ast$
cluster \citep{bec78,kra95,tan06}, the Quintuplet cluster
\citep{oku90,nag90}, and candidates in their vicinity \citep{nag93} to
use as probes of foreground H$_3^+$ (Table~\ref{tb1}). Spectroscopic
observations were carried out at the Subaru Telescope over several
nights between 2004 July 7 UT and 2004 Sep 26 UT, using the Infrared
Camera and Spectrograph (IRCS) \citep{tok98,kob00}. The data obtained
on 2001 Jun 16 UT, originally published in \citet{got02}, were
combined with the newer data. A summary of the observations is given
in Table~\ref{tb2}.

A curvature sensing adaptive optics (AO) system \citep{tak04} was used
whenever a wavefront reference star was available. Use of the AO
system significantly increased the throughput of starlight through the
narrow (0\farcs15) slit, which was necessary to attain high spectral
resolution ($\lambda / \Delta \lambda$ = 20,000; $\Delta v =$
15~km~s$^{-1}$). In order to subtract sky and telescope emission,
either the telescope itself or the tip-tilt mirror in the AO system
was nodded by about 3\arcsec~ along the slit so that the stellar
spectrum fell on two different parts of the camera. The slit position
was set along the east-west axis in all cases except when nearby faint
sources fell onto the slit, which would have hampered a clean
sky-subtraction.

The angle of cross-dispersing grating in IRCS was adjusted so that as
many H$_3^+$ lines as possible were observed in the time
available. The major absorption lines essential for the gas
diagnostics (the $R$(1,0), $R$(1,1)$^u$, $R$(1,1)$^l$, $Q$(1,0), and
$Q$(1,1) lines from the lowest (1,1) and (1,0) levels, and the
$R$(3,3)$^l$ and $R$(2,2)$^l$ lines from the higher (3,3) and (2,2)
levels) were covered in two grating settings. The $Q$(1,0) transition
at 3.9530~$\mu$m was covered by both of the grating settings. Details
of the spectral setup are found in \citet{got02}.

Standard stars with early spectral types were observed on the same
nights through similar airmasses as the Galactic center stars in order
to remove the telluric absorption features. Spectroscopic flat-fields
were obtained from a halogen lamp installed at the Cassegrain port in
front of the instrument's entrance window. The sky conditions were
fair with the seeing between 0\farcs45 and 0\farcs75 at $R$ band.

\begin{table*}
\tablewidth{\textwidth}
\begin{center}
\scriptsize
  \caption{Summary of observations.}\label{tb2}
  \begin{tabular}{lrrrllll}
\hline \hline
Source                & Exposure &\multicolumn{2}{c}{Grating$^a$}&Line Coverage                            &\multicolumn{2}{c}{Atmospheric Standard}  & UT Date     \\
                      &   [s]    & ECH   & XDP  &                                                             &   Name                 & Spe.    &             \\
\hline

GC~IRS~21        \dots&  1800    & 11050 &2800  &$R$(2,2)$^l$, $R$(1,1)$^u$, $R$(1,0), $Q$(1,1), $Q$(1,0)     & HR~7528                & B9.5~IV & 2004 Jul 28 \\
                      &  2400    &  5050 &3500  &$R$(3,3)$^l$, $R$(1,1)$^l$, $Q$(1,0)                         & HR~6378                & A2~V    & 2004 Sep 02 \\
                      &          &       &      &                                                             &                        &         &             \\

GC~IRS~3        \dots &  360     &  8350 &6100  &$R$(1,1)$^u$, $R$(1,0), $R$(1,1)$^l$,$Q$(1,1),$Q$(1,0)       & HR~7924 ($\alpha$ Cyg) & A2~Iae  & 2001 Jun 16 \\
                      &  1440    &  4400 &5200  &$R$(3,3)$^l$, $R$(1,1)$^l$                                   & HR~7924 ($\alpha$ Cyg) & A2~Iae  & 2001 Jun 16 \\

                      &  3120    & 11050 &2800  &$R$(2,2)$^l$, $R$(1,1)$^u$, $R$(1,0), $Q$(1,1), $Q$(1,0)     & HR~7001 ($\alpha$ Lyr) & A0~V    & 2004 Jul 07 \\
                      &  2400    &  5050 &3500  &$R$(3,3)$^l$, $R$(1,1)$^l$, $Q$(1,0)                         & HR~7001 ($\alpha$ Lyr) & A0~V    & 2004 Jul 08 \\

                      &  3000    & 11050 &2800  &$R$(2,2)$^l$, $R$(1,1)$^u$, $R$(1,0), $Q$(1,1), $Q$(1,0)     & HR~7121                & B2.5~V  & 2004 Sep 25 \\
                      &  1280    & 11050 &2800  &$R$(2,2)$^l$, $R$(1,1)$^u$, $R$(1,0), $Q$(1,1), $Q$(1,0)     & HR~7121                & B2.5~V  & 2004 Sep 26 \\
                      &          &       &      &                                                             &                        &         &             \\

GC~IRS~1W        \dots&  2560    &  5050 &3500  &$R$(3,3)$^l$, $R$(1,1)$^l$, $Q$(1,0)                         & HR~7001 ($\alpha$ Lyr) & A0~V    & 2004 Jul 27 \\
                      &  2560    & 11050 &2800  &$R$(2,2)$^l$, $R$(1,1)$^u$, $R$(1,0), $Q$(1,1), $Q$(1,0)     & HR~7528                & B9.5~IV & 2004 Jul 27 \\
                      &          &       &      &                                                             &                        &         &             \\

NHS~21           \dots&  3600    & 11050 &2800  &$R$(2,2)$^l$, $R$(1,1)$^u$, $R$(1,0), $Q$(1,1), $Q$(1,0)     & HR~7001 ($\alpha$ Lyg) & A0~V    & 2004 Jul 08 \\
                      &  3600    &  5050 &3500  &$R$(3,3)$^l$, $R$(1,1)$^l$, $Q$(1,0)                         & HR~6378                & A2~V    & 2004 Jul 08 \\
                      &          &       &      &                                                             &                        &         &             \\

NHS~22           \dots&  1200    & 11050 &2800  &$R$(2,2)$^l$, $R$(1,1)$^u$, $R$(1,0), $Q$(1,1), $Q$(1,0)     & HR~7121, ATRAN         & B2.5~V  & 2004 Sep 04 \\
                      &  1200    &  5050 &3500  &$R$(3,3)$^l$, $R$(1,1)$^l$, $Q$(1,0)                         & HR~7121, ATRAN         & B2.5~V  & 2004 Sep 04 \\
                      &          &       &      &                                                             &                        &         &             \\

NHS~42           \dots&  1920    & 11050 &2800  &$R$(2,2)$^l$, $R$(1,1)$^u$, $R$(1,0), $Q$(1,1), $Q$(1,0)     & HR~7924 ($\alpha$ Cyg) & A2~Iae  & 2004 Jul 27 \\
                      &  1920    &  5050 &3500  &$R$(3,3)$^l$, $R$(1,1)$^l$, $Q$(1,0)                         &                        &         & 2004 Jul 27 \\
                      &          &       &      &                                                             &                        &         &             \\

NHS~25           \dots&  1200    &  5050 &3500  &$R$(3,3)$^l$, $R$(1,1)$^l$, $Q$(1,0)                         & HR~7121                & B2.5~V  & 2004 Sep 04 \\
                      &  1200    & 11050 &2800  &$R$(2,2)$^l$, $R$(1,1)$^u$, $R$(1,0), $Q$(1,1), $Q$(1,0)     & HR~7121                & B2.5~V  & 2004 Sep 04 \\
                      &          &       &      &                                                             &                        &         &             \\

GCS~3-2       \dots   &  120     &  8350 &6100  &$R$(1,1)$^u$, $R$(1,0), $R$(1,1)$^l$, $Q$(1,1),$Q$(1,0)      & HR~7528                & B9.5~IV & 2001 Jun 16 \\
                      &  240     &  4400 &5200  &$R$(3,3)$^l$, $R$(1,1)$^l$, $Q$(1,0)                         & HR~7528                & B9.5~IV & 2001 Jun 16 \\

                      &  1344    & 11050 &2800  &$R$(2,2)$^l$, $R$(1,1)$^u$, $R$(1,0), $Q$(1,1), $Q$(1,0)     & HR~7001 ($\alpha$ Lyr) & A0~V    & 2004 Jul 07 \\
                      &  1344    &  5050 &3500  &$R$(3,3)$^l$, $R$(1,1)$^l$, $Q$(1,0)                         & HR~7001 ($\alpha$ Lyr) & A0~V    & 2004 Jul 07 \\

\hline
    \end{tabular}

\tablenotetext{a}{ECH and XDP denote the angles of echelle and
cross-dispersing gratings in the instrumental coding unit.}

\end{center}
\end{table*}

\section{Data Reduction}

The data were reduced using IRAF\footnote{IRAF is distributed by the
National Optical Astronomy Observatories, which are operated by the
Association of Universities for Research in Astronomy, Inc., under
cooperative agreement with the National Science Foundation.} and
custom-written IDL codes. The two dimensional spectrograms were
pair-subtracted and then averaged and the pixel responses were
normalized by the dark-subtracted flat-field images. Noisy pixels were
filtered out based on the pixel statistics, and interpolated prior to
spectral extraction. The one-dimensional spectra were obtained using
the IRAF aperture extraction package.

The removal of the telluric lines by dividing the data by the spectra
of the standard stars was handled by IDL codes that take into account
possible observational flaws, including slight shifts in the
wavelengths, differences in airmasses (and therefore depths of
telluric absorption lines), slightly different spectral resolutions,
fringing of the continua, and saw-toothed features due to the
different readout channels. In cases where the flaws were too extreme
to be easily corrected, a transmission spectrum calculated by ATRAN
\citep{lor92} was used as a substitute to remove the telluric
features. The signal-to-noise ratios of the spectra were calculated
from the noises on the continua near the H$_3^+$ absorption lines.

Wavelength calibration was achieved by matching the observed spectra
to the model transmission spectrum. The accuracy of the calibration is
normally better than one tenth of one pixel ($\delta v$ =
0.75~km~s$^{-1}$) when the signal-to-noise ratio of the observed
spectrum is sufficiently high.
The radial velocities of the absorption lines observed in the multiple
runs were corrected for the orbital motion of the Earth, converted to
$v_{\rm LSR}$ using the radial velocity package of IRAF, and added
together with appropriate weights according to the signal-to-noise
ratios of the spectra. The signal-to-noise ratios after adding up all
available data range 60 to 300.

\begin{figure*}
\begin{center}
\includegraphics[angle=0,width=.9\textwidth]{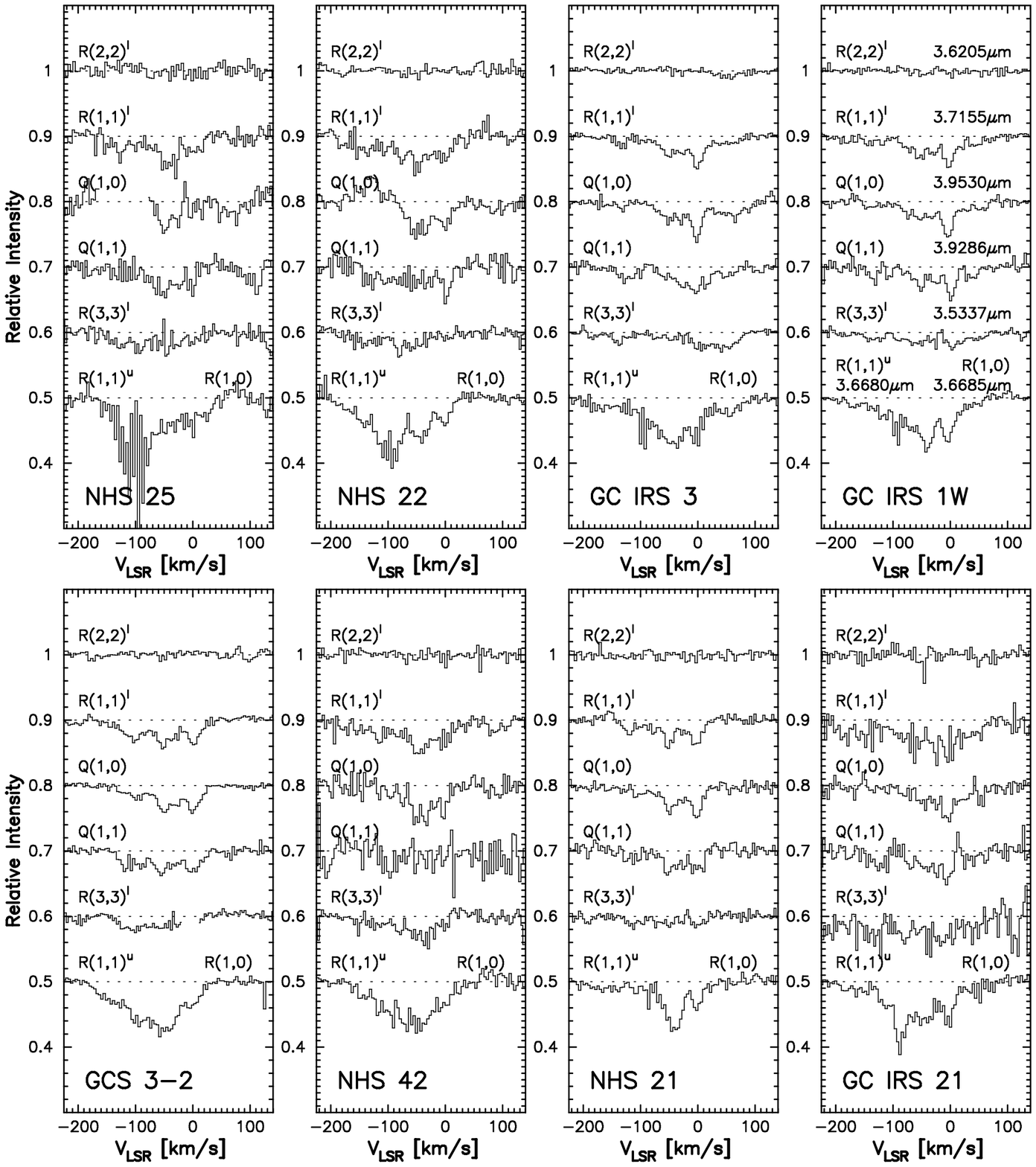}
\caption{Spectra of H$_3^+$ toward eight sources in the Central
Molecular Zone. The $R$(1,1)$^u$ and $R$(1,0) absorptions in the
bottom trace are separated by 35~km~s$^{-1}$ and the profiles
overlap. The wavelength interval near $Q$(1,1) is heavily interfered
by the telluric absorption of N$_2$O, causing the low
signal-to-noise ratios. The gap in the spectrum of NHS 25 in the
$Q$(1,0) line corresponds to a strong atomic emission line, probably
of Si II. \label{f1}}
\end{center}
\end{figure*}

The seven most relevant lines for this study are shown in
Fig.~\ref{f1}. For the Pistol Star, NHS 25 and possibly more weakly
for NHS 22, the $Q$(1,0) line at 3.9530 $\mu$m is overlapped at low
wavelength with an atomic emission line.  This wavelength region is
omitted for NHS 25 in Fig.~1. The atomic emission matches the Si II
lines at 3.9477~--~3.9499~$\mu$m compiled in the NIST table
\citep{ral07}.

\section{Results}

\subsection{Overview}

All five of the absorption lines ($R$(1,0), $R$(1,1)$^u$,
$R$(1,1)$^l$, $Q$(1,0), and $Q$(1,1)) that originate from the lowest
para and ortho rotational levels, ($J$, $K$) = (1,1) and (1,0), were
detected in all sight lines. In each sight line the line profiles of
these transitions are basically the same (Fig.~\ref{f1}), showing a
broad trough of absorption primarily at negative radial velocities,
and several sharp absorption minima, which we attribute to intervening
clouds in the foreground spiral arms, as did \citet{oka05} for
GCS~3-2.

The 3.5337~$\mu$m $R(3,3)^l$ line from the (3,3) metastable level was
detected in all targets. In nearly every case it occurs largely at
negative radial velocities. Its profile is broad and rather smooth,
with a width approaching 150~km~s$^{-1}$, and with no sharp features.
H$_3^+$ lines of such width have never been seen in the Galactic
disk. Moreover, searches for the $R(3,3)^l$ line in the disk along
sight lines where the H$_3^+$ lines from the lowest two levels have
clearly been observed \citep{geb96,mcc98,geb99,mcc02,ind07} have been
unsuccessful \citep{oka05}. For these reasons and following
\citet{oka05}, we attribute the entire $R$(3,3)$^l$ absorption to the
gas in the CMZ. The $R(3,3)^l$ line profile mimics the broad trough in
the $J$~=~1 $\mathrm{H}_{3}^{+}$ absorption line profiles. We use this
resemblance to differentiate between the absorptions in the CMZ and
those in the spiral arms.

\begin{figure}
\begin{center}
\includegraphics[angle=0,width=.32\textwidth]{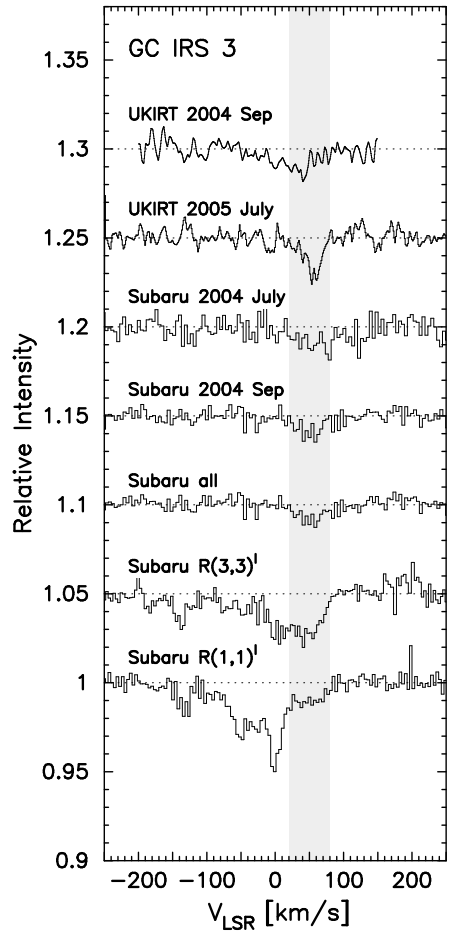}
\caption{Spectra of the $R$(2,2)$^l$ line at 3.6205~$\mu$m toward
GC~IRS~3 obtained at Subaru and UKIRT. The $R$(3,3)$^l$ and
$R$(1,1)$^l$ lines observed at Subaru are also shown. \label{f2}}
\end{center}
\end{figure}

The $R(2,2)^l$ line at 3.6205~$\mu$m was detected in one sight line
only, that toward GC~IRS~3, and only at velocities near
$+$50~km~s$^{-1}$. Because this is the first time that this line has
been detected and because absorption by it is only 1 -- 2~\%, we have
carefully examined its reality. Two telluric CH$_{4}$ lines line at
2762.48~cm$ ^{-1}$ (3.61994~$\mu$m) and 2761.36~cm$^{-1}$
(3.62141~$\mu$m) with peak absorption strengths of 20 -- 30~\%\
overlap the $R$(2,2)$^l$ line; incomplete cancellation of them could
cause absorption features near the observed feature. However, the
wavelength match is not precise and moreover, the width of the
$R$(2,2)$^l$ line (40 -- 60~km~s$^{-1}$) is twice that of each of the
telluric lines, which are nearly unresolved. As a further check we
obtained independent observations of GC~IRS~3 from the United Kingdom
Infrared Telescope (UKIRT) in September 2004 and July 2005, using the
echelle in the facility spectrograph CGS4. Although somewhat different
in detail, the UKIRT observations consistently show a 1 -- 2~\%
absorption dip at the same wavelength as the Subaru data. Thus the
detection of this line toward GC~IRS~3 is genuine.  All observed
spectra of the $R$(2,2)$^l$ line in GC~IRS~3 are shown in
Fig. \ref{f2} together with the $R$(3,3)$^l$ and $R$(1,1)$^l$
absorption lines.

In contrast to the Galactic disk where the detectable presence of
H$_3^+$ is limited to relatively few sight lines and their narrow
angular proximity, every sight line observed toward the CMZ has
abundant H$_3^+$. Although the targets presented here are within 30~pc
of Sgr~A$^\ast$ in projected distance, the lengths of all of the
absorbing columns of H$_3^+$ are very large (significantly larger than
this separation) as discussed below (see also \citet{oka05}) and the
observed velocity dispersion is very large, with absorption occurring
from $-$150~km~s$^{-1}$ to $+$20~km~s$^{-1}$.  This range
approximately matches the entire span of the radial velocity of the
gas in the CMZ on the front-side of the center \citep{saw04}. These
characteristics suggest that the gas containing the observed H$_3^+$
is not located in well-separated entities, but is continuous with a
large volume filling factor. The large column densities of H$_3^+$ in
all sight lines are most obvious in the bottom traces of Fig.~\ref{f1}
where two of the strong H$_3^+$ spectral lines from the (1,1) and
(1,0) levels overlap and yield a blended feature of large equivalent
width.

\subsection{$\mathrm{H}_{3}^{+}$ equivalent widths and column densities}

H$_3^+$ lines from the $J$ = 1 levels are composites of absorption in
the CMZ and in foreground spiral arms. \citet{oka05} separated the two
components of the $R$(1,1)$^l$ absorption profile in GCS~3-2 by using
the 2.3~$\mu$m \mbox{v~=~2~--~0} $R$(1) line of CO which is composed
of sharp absorption features at the radial velocities
($-$50~km~s$^{-1}$, $-$30~km~s$^{-1}$ and 0~km~s$^{-1}$) of foreground
spiral arms with little or no CMZ component. The sharp features in the
H$_3^+$ absorption profile, which mimic the CO spectrum, were
subtracted from the $R$(1,1)$^l$ line and the remainder was ascribed
to H$_3^+$ in the CMZ \citep[see Fig.~4 of ][]{oka05}.

\begin{figure*}[bth]
\begin{center}
\includegraphics[angle=-90,width=.9\textwidth]{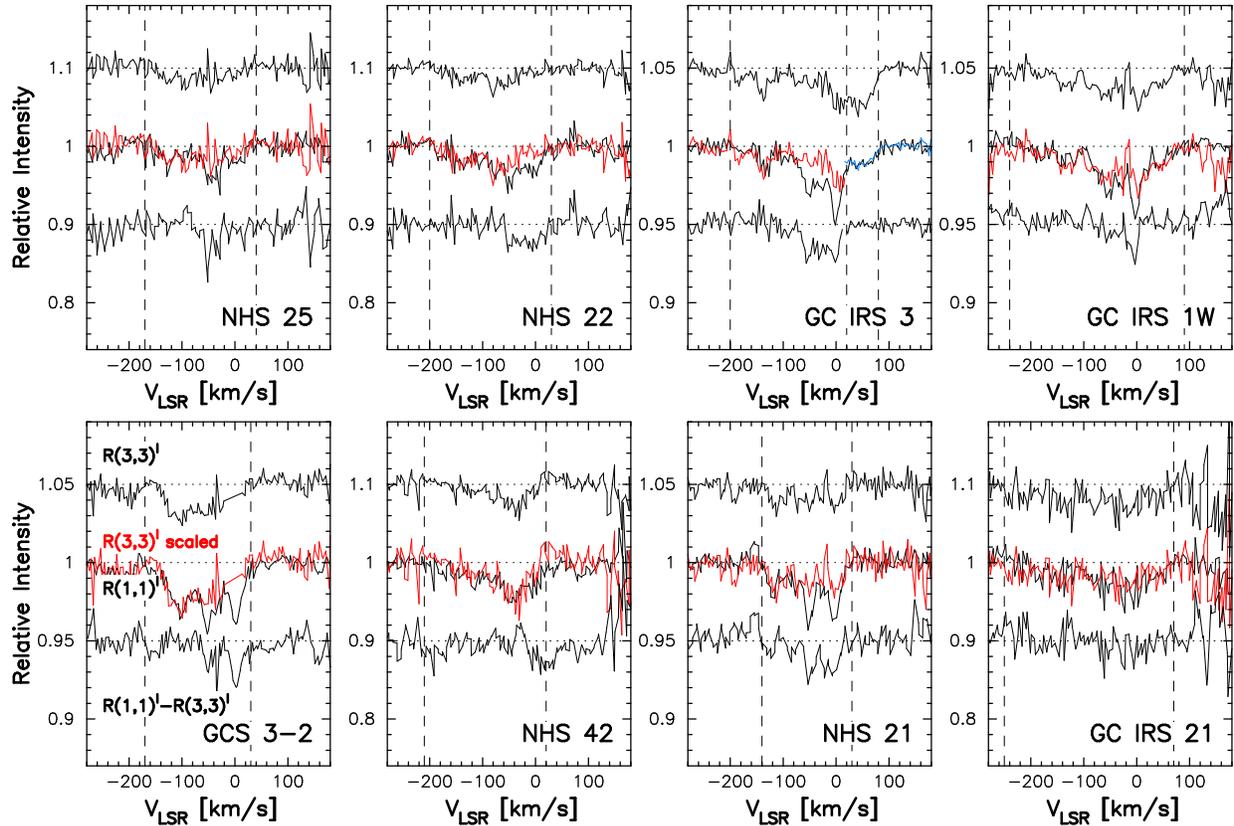}
\caption{Scaling the $R$(3,3)$^l$ absorption (top trace in each
panel) to match the absorption trough of the $R$(1,1)$^l$ absorption
(middle trace in black). The scaled $R$(3,3)$^l$ is overlaid in red
with $R$(1,1)$^l$, and the difference spectrum is shown in the
bottom trace. Vertical lines indicate the range of velocities where
the profiles were scaled. For GC~IRS~3, the scaling is done
separately in the two velocity ranges, $-200$--20~km~s$^{-1}$ (red)
and 20--80~km~s$^{-1}$ (blue). \label{f3}}
\end{center}
\end{figure*}

We found that this method was not effectively applicable for the
spectra in Fig.~\ref{f1} because of the lower resolution of the IRCS
(by a factor of 3) compared to the Phoenix Spectrometer of the Gemini
South Observatory, and the low signal-to-noise ratios of spectra
toward targets fainter than GCS~3-2. We used an alternative method for
separation, in which we assume for each star that the velocity profile
of the trough in the $J$~=~1 lines is a scaled version of the
$R$(3,3)$^l$ profile. The method is depicted in Fig.~\ref{f3}. It
crudely assumes uniform physical conditions in the absorbing clouds
along the line of sight. We used it to separate the CMZ and spiral arm
components of the $Q$(1,0) line as well as the $R$(1,1)$^l$ line.  We
ignore the detailed velocity profiles of the observed spectra, and
list integrated absorptions of the $R$(1,1)$^l$ line and their ratios
in the CMZ to the total (CMZ plus spiral arms). This ratio is also the
ratio of column densities. We tested the method for GCS~3-2 for which
\citet{oka05} used the high resolution spectrum obtained by Phoenix at
the Gemini South Observatory and reported the ratio ($N(1,1)_{\rm
GC}/N(1,1)_{\rm Total} = 0.68$) by a different method of analysis
using CO spectrum; the agreement is satisfactory. The interference of
the $Q$(1,0) line by the atomic emission for NHS~25 mentioned in
Section~3 does not seriously hamper the measurement since the scaling
of the $R$(3,3)$^l$ line to the $Q$(1,0) line is not much affected.
The ratio of the equivalent width of the CMZ component to the total
ranges from 0.5 to 0.9 for the eight sight lines.

The sight line toward GC~IRS~3 is exceptional in that it passes
through two different categories of gas. In addition to the gas with
mostly negative velocity common to all eight targets, it shows
absorption at $+$50~km~s$^{-1}$. This absorption is unique in two
other ways: (1) it is the only gas in which the $R$(2,2)$^l$ line is
observable (indicating that the gas producing this velocity feature is
higher density than the gas seen at other velocities) and (2) it is
the only gas for which the column density of the (3,3) metastable
level is definitely higher than in the (1,1) ground level. Therefore
we consider the $+$50~km~s$^{-1}$ absorption separately. Measured
equivalent widths of all spectral lines are listed in
Table~\ref{tb3}. For those transitions starting from the (1,1) level,
both total equivalent widths and equivalent widths for the part of the
absorption originating in the CMZ are provided.

The H$_3^+$ column densities in the (1,0), (1,1), (3,3), and (2,2)
levels were calculated from the equivalent widths of the $Q$(1,0)
$R$(1,1)$^l$, $R$(3,3)$^l$, and $R$(2,2)$^l$ lines, respectively, as
in \citet{geb99}, from the equation $ W_\lambda = (8
\pi^3/3h\nu)N_{\rm level} |{\boldmath \mu}|^2$, using the theoretical
line strengths $|{\boldmath \mu}|^2$ provided by J. K. G. Watson
\citep[also calculable from the list of][]{nea96}. The results are
given in Table~\ref{tb4}. The column density corresponding to the 50
km~s$^{-1}$ absorption toward GC~IRS~3 is listed separately at the
bottom of the table. The total column densities $N$(H$_3^+$)$_{\rm
total}$ are simply the sum of column densities of the (1,1), (1,0),
(3,3), and (2,2) levels since modeling of the thermalization
\citep{oka04} shows that they are the only levels significantly
populated at the temperatures and densities of the region.

For broad spectral lines such as those treated in this paper,
uncertainties in the equivalent widths result mainly from the
uncertainties in defining the continuum level rather than the standard
deviations of individual data point. The listed uncertainties were
estimated by multiplying the standard deviation of the continuum level
by the width of the relevant velocity range.


\subsection{Temperature and Density}

The column densities given in Table~\ref{tb4} were used to determine
the temperatures, $T$, and densities, $n$, of the gas based on the
steady state calculation by \citet{oka04}. Their diagram, in which the
values of the population ratios $n$(3,3)/$n$(1,1) and
$n$(3,3)/$n$(2,2) are plotted as functions of $T$ and $n$ (their
Fig.~4), is inverted in our Fig.~\ref{f4} so that the observed
population ratios can be used directly to determine $T$ and $n$. The
population ratios toward GCS~3-2 from the velocity resolved data by
\citet{oka05} are included in the figure for comparison. Most of the
clouds in the Galactic center fall in the narrow parameter region of
the $T-n$ diagram, 200~K$<T<$~300~K and $n$ $\leq$ 50--200~cm$^{-3}$,
with the sole exception being the $+$50~km~s$^{-1}$ component toward
GC~IRS~3 for which a higher temperature of $\sim$ 400~K and a density
of 200~cm$^{-3}$ were found.

The upper limits to the number densities in Table~\ref{tb4}  
are based on the upper 
limits to the $R$(2,2)$^l$ line equivalent widths. Lower
limits may be roughly estimated from the total extinction of $A_V$
$\sim$ 30~mag reported by \citet{cot00} toward the Quintuplet and
Central Clusters and the estimate by \citet{whi97} that the visual
extinction of the diffuse interstellar medium toward the GC is
$\sim$ 20 mag, that is, $E$($B - V$) $\sim$ 6.5. Ascribing all of
the latter to the CMZ and using the standard conversion factor from
$E(B - V)$ to $N$(H) \citep{boh78} gives
$N$(H$_2$)~$\sim$~2~$\times$~10$^{22}$~cm$^{-2}$. Assuming $L$ = 100
pc yields $n$(H$_2$) $\sim$~60~cm$^{-3}$. This shows that the gas
density is not much lower than the upper limit. Indeed the very
existence of observable $\mathrm{H}_{3}^{+}$ also argues against
much lower densities, in which conditions it would be difficult for
the H$_2$, which is needed to produce H$_3^+$, to exist.

\begin{table*}
\begin{center}
\tablewidth{\textwidth}
\tiny
  \caption{Equivalent widths of absorption lines.\label{tb3}}

\begin{tabular}{l cc c cc cccc c}
\hline \hline
       &\multicolumn{2}{c}{Velocity Range\tablenotemark{a}}& $W_\lambda$ Total\tablenotemark{b} & \multicolumn{2}{c}{Scaling Factor \tablenotemark{c}}
       & \multicolumn{4}{c}{$W_{\lambda}$ in Galactic Center\tablenotemark{d}} & \\
       &\multicolumn{2}{c}{---------------------------}
       &  
       & \multicolumn{2}{c}{--------------------------------}
       & \multicolumn{4}{c}{--------------------------------------------------------------} &      \\
Source        & $v_1$ &  $v_2$
              & $R$(1,1)$^l$
              & $R$(3,3)$^l$ & $R$(3,3)$^l$
              & $R$(1,1)$^l$ & $Q$(1,0) & $R$(3,3)$^l$ & $R$(2,2)$^l$
              & $N(1,1)_{\rm GC}/N(1,1)_{\rm Total}$\tablenotemark{e} \\

              &[km~s$^{-1}$]  &[km~s$^{-1}$]
              &[10$^{-5}\mu$m]
              & $\rightarrow R$(1,1)$^l$    & $\rightarrow Q$(1,0)
              &[10$^{-5}\mu$m]&[10$^{-5}\mu$m]&[10$^{-5}\mu$m]&[10$^{-5}\mu$m]
              & \\

\hline

GC~IRS~21\dots &  $-$250 &   70 &  6.7$\pm$ 2.7 & 0.90 & 0.55 &  6.1$\pm$ 2.4 &  3.7$\pm$ 1.0 &  6.8$\pm$ 3.4 & $<$ 2.2 & 0.91$\pm$0.40 \\
GC~IRS~3 \dots &  $-$200 &   20 &  4.0$\pm$ 0.5 & 1.00 & 0.60 &  2.4$\pm$ 0.5 &  1.4$\pm$ 0.3 &  2.4$\pm$ 0.5 & $<$ 0.4 & 0.59$\pm$0.12 \\
GC~IRS~1W\dots &  $-$240 &   90 &  4.5$\pm$ 0.8 & 1.20 & 1.00 &  3.9$\pm$ 1.0 &  3.3$\pm$ 1.0 &  3.3$\pm$ 0.8 & $<$ 0.9 & 0.88$\pm$0.18 \\
NHS~21   \dots &  $-$140 &   30 &  3.6$\pm$ 0.5 & 1.25 & 1.00 &  2.0$\pm$ 0.6 &  1.6$\pm$ 0.6 &  1.6$\pm$ 0.4 & $<$ 0.6 & 0.54$\pm$0.13 \\
NHS~22   \dots &  $-$200 &   30 &  5.9$\pm$ 1.2 & 1.35 & 1.20 &  3.7$\pm$ 1.7 &  3.3$\pm$ 1.8 &  2.7$\pm$ 0.7 & $<$ 1.0 & 0.63$\pm$0.21 \\
NHS~42   \dots &  $-$210 &   20 &  6.4$\pm$ 1.1 & 1.60 & 1.45 &  3.9$\pm$ 1.8 &  3.5$\pm$ 2.5 &  2.4$\pm$ 1.1 & $<$ 1.1 & 0.61$\pm$0.17 \\
NHS~25   \dots &  $-$170 &   40 &  4.2$\pm$ 1.3 & 1.20 & 0.80 &  2.5$\pm$ 1.5 &  1.7$\pm$ 1.3 &  2.1$\pm$ 1.5 & $<$ 1.0 & 0.60$\pm$0.31 \\
GCS~3-2  \dots &  $-$170 &   30 &  5.5$\pm$ 0.4 & 1.35 & 0.70 &  3.7$\pm$ 0.5 &  1.9$\pm$ 0.2 &  2.7$\pm$ 0.5 & $<$ 0.8 & 0.68$\pm$0.07 \\

\hline
GC~IRS~3~(50~km~s$^{-1}$)\tablenotemark{f}\dots &     20 &   80 &  0.7$\pm$ 0.1 & 0.50 & 0.65 &  0.7$\pm$ 0.1 &  0.9$\pm$ 0.1 &  1.4$\pm$ 0.1 & 0.5 $\pm$ 0.2 & 0.93$\pm$0.15 \\
               &         &      &               &      &      &               &               &               &         &               \\
\hline
   \end{tabular}

\tablenotetext{a}{Equivalent widths are measured by integrating
absorption intensity in the range between $v_1$  and $v_2$. }

\tablenotetext{b}{The total equivalent width of the $R$(1,1)$^l$
line.}

\tablenotetext{c}{Scaling factor to match the $R$(3,3)$^l$
absorption lines to the troughs of the $R$(1,1)$^l$ lines(see
Fig.~\ref{f3}). }

\tablenotetext{d}{Equivalent widths of H$_3^+$ absorption troughs
due to H$_3^+$ in the CMZ.}

\tablenotetext{e}{Ratio of equivalent width in the CMZ to the total
equivalent width.}

\tablenotetext{f}{Equivalent width of GC~IRS~3 in the
20-80~km~s$^{-1}$ range where the $R$(2,2)$^l$ absorption is
detected.}


\normalsize
\end{center}
\end{table*}


\section{Discussion} 

\subsection{The Ionization Rate}

An estimate of the ionization rate, $\zeta$, in the Galactic center
is the most important result of these $\mathrm{H}_{3}^{+}$
observations. To avoid confusion, we use $\zeta$(H) and
$\zeta$(H$_2$) for the ionization rates of H and H$_2$,
respectively, in the Galactic disk where ionization is mainly due to
cosmic-rays. We use $\zeta$ for the rate in the CMZ where the effect
of X-rays and FUV radiation may be significant. $\zeta$(H) and
$\zeta$(H$_2$) are related approximately as $\zeta$(H$_2$) =
2$\zeta$(H) \citep{dal06}.

Early studies of the cosmic ray ionization rate were made in order
to understand the physical state of the interstellar medium and its
heating mechanisms \citep{hay61,spi68,fie69}. \citet{spi68} reported
a minimum value of
\emph{$\zeta_f$}~=~6.8~$\times$~10$^{-18}$~s$^{-1}$ from an
extrapolation of the cosmic ray spectrum observed at Earth, and an
upper limit of 1.2 $\times$ 10$^{-15}$ s$^{-1}$ calculated from an
assumed supernova frequency. Their \emph{$\zeta_f$} is the total
ionization rate for atomic hydrogen including the effect of
secondary electrons, written here as \emph{$\zeta$}(H). The most
recent measurement of \emph{$\zeta$}(H), reported by \citet{web98}
using data from the \emph{Voyager} and \emph{Pioneer} spacecraft at
distances up to 60 AU from the Sun, is
\emph{$\zeta$}(H)~$\geq$~(3--4)~$\times~10^{-17}$~s$^{-1}$, which
should replace the lower limit used by \citet{spi68}.

With the advent of the {\it Copernicus} satellite and its
far-ultraviolet spectrometer, which allowed FUV studies of chemical
abundances in the diffuse ISM \citep[for a review; ][]{spi75},
column densities of molecules and radicals such as HD and OH could
be measured.  The reaction sequences leading to the creation of
these two molecules start from H$^+$ and are thus useful for
estimating \emph{$\zeta$}(H). Values of \emph{$\zeta$}(H) determined
in this way \citep{odo74,bla77,har78} are mostly a few times
10$^{-17}$ s$^{-1}$, comparable to the lower limit given above. The
comprehensive model of the diffuse interstellar medium by
\citet{dis86} suggested a somewhat higher value of
\emph{$\zeta$}(H)~$\approx$~7~$\times$~10$^{-17}$~s$^{-1}$. However
they used an H$_3^+$ dissociative recombination rate constant one
thousand times lower than the current value which was popular in
some circles at the time. They noted that if the thousand times
higher constant were to be used, the value of \emph{$\zeta$}(H)
would need to be increased by a factor of $\sim$ 10.


The formation and destruction mechanisms for H$_3^+$ are the simplest
among all astrophysical probes of the ionization rate, and this makes
H$_3^+$, when detectable, a powerful tool for measuring the ionization
rate. While the H$_3^+$ column densities observed in dense clouds,
\mbox{(0.4~--~2.3)~$\times$~10$^{14}$~cm$^{-2}$}, are consistent with
\emph{$\zeta$}(H$_2$) on the order of 3~$\times$~10$^{-17}$~s$^{-1}$
\citep{geb96, mcc99}, the surprisingly high $N$(H$_3^+$) observed in
diffuse clouds, comparable to those in dense clouds in spite of their
10 times smaller visual extinction \citep{mcc98, geb99, mcc02},
require an order of magnitude higher \emph{$\zeta$}(H$_2$). Crucial to
this conclusion are the laboratory measurements of the dissociative
recombination rate constant $k_e$ in Eq.~(1)
\citep{ama88,lar93,lar00,mcc03,mcc04}.

Together, \citet{mcc02} and \citet{ind07} have carried out a survey
of H$_3^+$ in the diffuse interstellar medium toward 29 nearby
stars. H$_3^+$ was detected in 14 of these sight lines with column
densities of (0.6~--~6.5)~$\times$~10$^{14}$~cm$^{-2}$. For these
diffuse clouds the product $\zeta$$L$ is in the range
(0.5~--~4.4)~$\times$~10$^4$~cm~s$^{-1}$. Separating $\zeta$ and $L$
is not straightforward. Based on various methods of estimating $L$,
which gave values of 2.2~-~31 pc for different clouds, the cosmic
ray ionization rates were found to be in the range
\emph{$\zeta$}(H$_2$)~=~(1.2~--~7.4)~$\times$~10$^{-16}$~s$^{-1}$
with an average of 5 $\times $10$^{-16}$~s$^{-1}$. Note that the
primary ionization rate of H, $\zeta$$_p$, used by \citet{ind07} is
related to $\zeta$(H$_2$) by 2.3$\zeta$$_p$ = $\zeta$(H$_2$)
\citep{gla74}. This has established that the cosmic ray ionization
rate in diffuse clouds is an order of magnitude higher than in dense
clouds. This result might be understandable in view of attenuation
of soft components of cosmic rays in dense clouds \citep{cra78}, if
a large flux of low energy cosmic rays is ubiquitous in the Galaxy
\citep{mcc03}, or if higher energy cosmic rays are trapped by
Alfv\'{e}n waves in diffuse clouds \citep{pad05}. The discrepancy
between the high \emph{$\zeta$}(H$_2$) and previous low
\emph{$\zeta$}(H) values determined from HD and OH has been
explained by \citet{lis03} as due to neutralization of H$^+$ by
charge exchange with small grains which had not been considered in
the previous studies, in which the neutralization of H$^+$ was
assumed to be due to much slower radiative recombination with
electrons.

As seen in Table~\ref{tb3}, the total column density of H$_3^+$ in
the CMZ toward the eight observed stars ranges over (1.8 -- 6.1)
$\times$ 10$^{15}$ cm$^{-2}$, more than an order of magnitude higher
than the column densities in diffuse clouds in the Galactic disk,
suggesting even higher values of $\zeta$ and $L$. Since we are
interested in setting the minimum value of both $\zeta$ and $L$, we
use the maximum value of $f$~=~1 and the minimum value of $R_X$ = 3
\citep{sod95, ari96} in Eq.~(1) and obtain the numerical expression,

\begin{equation}
(\zeta L)_{\rm min} = 7.4 \times 10^{-11} {\rm cm}^3 {\rm s}^{-1}
N({\rm H}_3^+)_{\rm total}.
\end{equation}

\noindent To obtain Eq.~(2), we have used
$k_e$~=~7.7~$\times$~10$^{-8}$~cm$^3$~s$^{-1}$ calculated for $T$ =
250 K from the temperature dependence reported by \citet{mcc04}
assuming the same electron temperature as the gas temperature, which
may be justified in view of the fast thermalization of electrons
with molecules and atoms. We used the C to H ratio in the solar
vicinity of 1.6 $\times$ 10$^{-4}$ given by \cite{sof04}, which is
close to the value of 1.4 $\times$ $10^{-4}$, determined directly
from infrared spectra of CO and H$_2$ toward an embedded young star
NGC~2024~IRS~2 \citep{lac94}, and $1.3 \times 10^{-4}$, found by
\citet{usu08} with the same technique but toward NGC~7538~IRS~1 in
the outer Galaxy. The observed $N$(H$_3^+$)$_{\rm total}$ yield
lower limits for $\zeta$$L$, ($\zeta$$L$)$_{\rm min}$ = (1.3 -- 4.5)
$\times$ 10$^5$ cm s$^{-1}$. Although lower limits, 
the values of ($\zeta$$L$)$_{\rm min}$ are
1000 times higher than values of
$\zeta$$L$ determined for dense clouds \citep{mcc99} and 10 times
higher than values for diffuse clouds in the Galactic disk
\citep{ind07}. We cannot cleanly separate $\zeta$ and $L$ based on 
current understanding of the CMZ, but discuss two extreme cases in the
following sections.

\begin{table*}
\begin{center}
\tablewidth{\textwidth}
\tiny

\caption{Column densities of H$_3^+$ in the Galactic center.\label{tb4}}

\begin{tabular}{l cccc cc cc c cc}
\hline \hline
       &\multicolumn{4}{c}{$N({\rm H_3^+})_{\rm Level}$\tablenotemark{a}}
       &\multicolumn{2}{c}{Relative Population\tablenotemark{a}}
       &
       &
       & & \\

       &\multicolumn{4}{c}{-----------------------------------------------------------------------}
       &\multicolumn{2}{c}{--------------------------------------------}
       & &
       &
       & & \\
       & $N(1,1)$ &$N(1,0)$ &$N(3,3)$ &$N(2,2)$
       & $N(3,3)/N(1,1)$    &$N(3,3)/N(2,2)$     
       & $n(\rm{H_2})$  & $T(\rm{H_2})$
       & $N({\rm H_3^+})_{\rm Total}$
       & $(\zeta L)_{\rm min}$     & $L$ \tablenotemark{b}\\
Source &[10$^{14}$cm$^{-2}$]&[10$^{14}$cm$^{-2}$]&[10$^{14}$cm$^{-2}$]&[10$^{14}$cm$^{-2}$]
       & &
       & [cm$^{-3}$]                             & [K]
       & [10$^{14}$cm$^{-2}$]
       & [$10^3$cm~s$^{-1}$]                            & [pc]\\

\hline


GC~IRS~21\dots & 28.1$\pm$12.3 &  9.0$\pm$ 4.4 & 24.2$\pm$12.1 & $<$ 8.3 & 0.86$\pm$0.57 & $>$2.93 & $<$ 125 & 150--450  &   61.3$\pm$  17.8 &      451$\pm$     131 &  146$\pm$  43\\
GC~IRS~3 \dots & 10.8$\pm$ 2.1 &  3.4$\pm$ 1.3 &  8.4$\pm$ 1.9 & $<$ 1.6 & 0.78$\pm$0.23 & $>$5.36 & $<$ 50  & 225--400  &   22.5$\pm$   3.2 &      166$\pm$      23 &   54$\pm$   8\\
GC~IRS~1W\dots & 18.1$\pm$ 3.8 &  7.9$\pm$ 2.4 & 11.7$\pm$ 3.0 & $<$ 3.3 & 0.65$\pm$0.21 & $>$3.53 & $<$ 75  & 200--300  &   37.6$\pm$   5.4 &      277$\pm$      40 &   90$\pm$  13\\
NHS~21   \dots &  8.9$\pm$ 2.2 &  3.7$\pm$ 1.5 &  5.6$\pm$ 1.4 & $<$ 2.3 & 0.62$\pm$0.22 & $>$2.44 & $<$ 125 & 175--275  &   18.2$\pm$   3.1 &      134$\pm$      22 &   44$\pm$   7\\
NHS~22   \dots & 16.9$\pm$ 5.6 &  7.9$\pm$ 3.5 &  9.7$\pm$ 2.7 & $<$ 3.6 & 0.57$\pm$0.25 & $>$2.73 & $<$ 100 & 150--275  &   34.5$\pm$   7.2 &      254$\pm$      53 &   82$\pm$  17\\
NHS~42   \dots & 17.7$\pm$ 5.1 &  8.4$\pm$ 4.1 &  8.6$\pm$ 4.1 & $<$ 4.0 & 0.49$\pm$0.27 & $>$2.13 & $<$ 100 & 125--250  &   34.6$\pm$   7.7 &      255$\pm$      57 &   83$\pm$  18\\
NHS~25   \dots & 11.4$\pm$ 5.9 &  4.0$\pm$ 3.8 &  7.4$\pm$ 5.3 & $<$ 3.9 & 0.65$\pm$0.57 & $>$1.89 & $<$ 175 & 100--300  &   22.8$\pm$   8.8 &      168$\pm$      64 &   54$\pm$  21\\
GCS~3-2  \dots & 17.0$\pm$ 1.7 &  4.6$\pm$ 0.8 &  9.8$\pm$ 1.6 & $<$ 3.0 & 0.57$\pm$0.11 & $>$3.21 & $<$ 80  & 200--250  &   31.4$\pm$   2.5 &      231$\pm$      18 &   75$\pm$   6\\

\hline

GC~IRS~3~(50~km~s$^{-1}$)\dots &  3.2$\pm$ 0.5 &  2.1$\pm$ 0.4 &  4.9$\pm$ 0.5 & 1.9$\pm$ 0.6 & 1.55$\pm$0.29& 2.61$\pm$0.50& 150--350    & 350--500  &  10.2$\pm$    0.8 &       75$\pm$       6 &   24$\pm$   2\\

\hline
   \end{tabular}

\tablenotetext{a}{All column densities are for the CMZ only.}
\tablenotetext{b}{Using $\zeta$ = 1 $\times$ 10$^{-15}$~s$^{-1}$.}

\normalsize
\end{center}
\end{table*}


\subsection{The Case of Large $L$ in the CMZ}

\indent \citet{oka05} favored a large $L$ and hence a high volume
filling factor for the newly found warm and diffuse molecular gas.
This was not only because the $R$(3,3)$^l$ metastable lines with
Large velocity widths and similar velocity profiles had been detected 
along all eight lines of sight toward sources scattered over 30~pc in 
projected distance from Sgr~A$^\ast$, but also because
the deduced $T$ and $n$ of the gas were similar in all cases. The
latter suggests that the absorbing diffuse material is continuous
rather than contained in several separate entities. Further support for
this interpretation is that the range of H$_3^+$ absorption velocities 
in each sight line spans the entire range velocities observed at 
radio and millimeter wavelengths with large apertures and 
interpreted as coming from the front side of the CMZ.
Extending spectroscopy of H$_3^+$ over more widely spaced
sight lines in the CMZ will constitute a more definitive test of the
filling factor. A long path length may also be favored over high
ionization, in that there are limits to the value of $\zeta$ for 
efficient production of H$_3^+$ as discussed in the next subsection.

$L$ is unlikely to be significantly larger than the radius of the
CMZ, which for the region of interest is $\sim$ 130 pc, as estimated
using Fig.~11b of \citet{saw04}. Setting $L$~=~100 pc, close to this
maximum permissible value, the aforementioned range of values of
($\zeta$$L$)$_{\rm min}$ yields the lower limits $\zeta$$_{min}$ =
(0.4 -- 1.5)~$\times$~10$^{-15}$~s$^{-1}$ for the various sight
lines. For lower values of $f$ and higher values of $R_X$,
$\zeta$$_{min}$ will be higher.

The large extent of the warm and diffuse molecular gas postulated
above conflicts with the previous concept of the gas in the CMZ,
represented, for example, in Fig.~9 of \citet{laz98} where the
ultra-hot X-ray emitting plasma fills the CMZ. The warm and diffuse
gas revealed in this paper cannot coexist with the ultra-hot plasma
because the plasma electrons will be immediately cooled and
molecules will be destroyed. Thus in the above model the volume
filling factor of the ultra-hot gas would be reduced by a large
factor.

The large extent of the ultra-hot plasma gas in the CMZ was proposed
because of observations of extensive X-ray emission near the GC
\citep[e.g.,][]{koy89,yam93,koy96}. \citet{sun93} and \citet{mar93}
observed similar X-ray emission, but interpreted it differently in
view of the difficulty of confining such a hot gas by the
gravitational potential of the Galactic center, and the
extraordinarily large energy input required to maintain the high
temperature. With the advent of the \emph{Chandra} X-ray satellite,
a great many X-ray point sources have been resolved in the GC. These
have been interpreted as explaining all \citep{wan02}, or a
significant fraction \citep{mun03}, of the observed intense X-rays
as being due to stellar sources such as cataclysmic variables and
young stellar objects. Additional observational reports also argue
against the ultra-hot plasma based on the close correlation between
the stellar distribution and the observed intensities of the X-rays
in the GC \citep [e.g.,][]{war06} and in the Galactic ridge \citep
[e.g.,][]{rev06}.  However, interpretation of the X-ray emission
remains controversial \citep[e.g.,][]{mun04,moN06,koy07}.

Yet another category of high temperature gas observed in the CMZ is
the hot gas observed in recombination lines and other phenomena.
\citet{laz98} studied the $\lambda^2$-dependent hyper-strong
radio-wave scattering and free-free radio emissions and absorptions
and interpreted the data as evidence for \emph{T$_e$} $\sim 10^6$ K
and \emph{n$_e$} $\sim 10$ cm$^{-3}$ gas with nearly 100\% surface
filling factor toward the CMZ \citep[see][for a different
interpretation]{gol06}. \citet{rod05} have observed fine structure and
recombination lines toward a sample of 18 sources and reported even
higher electron densities of 30 -- 100~cm$^{-3}$. \citet{laz98}
inferred that these regions were at interfaces between molecular
clouds and the X-ray plasma and proposed (see their Fig.~9) that the
high density molecular clouds with a volume filling factor of
perhaps \emph{f$_{\rm v}$} $\sim$ 0.1 are surrounded by the scattering
electron gas, and the rest of the space is filled with X-ray emitting
plasma. Our observations and analysis \citep[see also][]{oka05}
contradict such a picture \citep{bol06}.

\begin{figure*}
\begin{center}
\includegraphics[angle=-90,width=.8\textwidth]{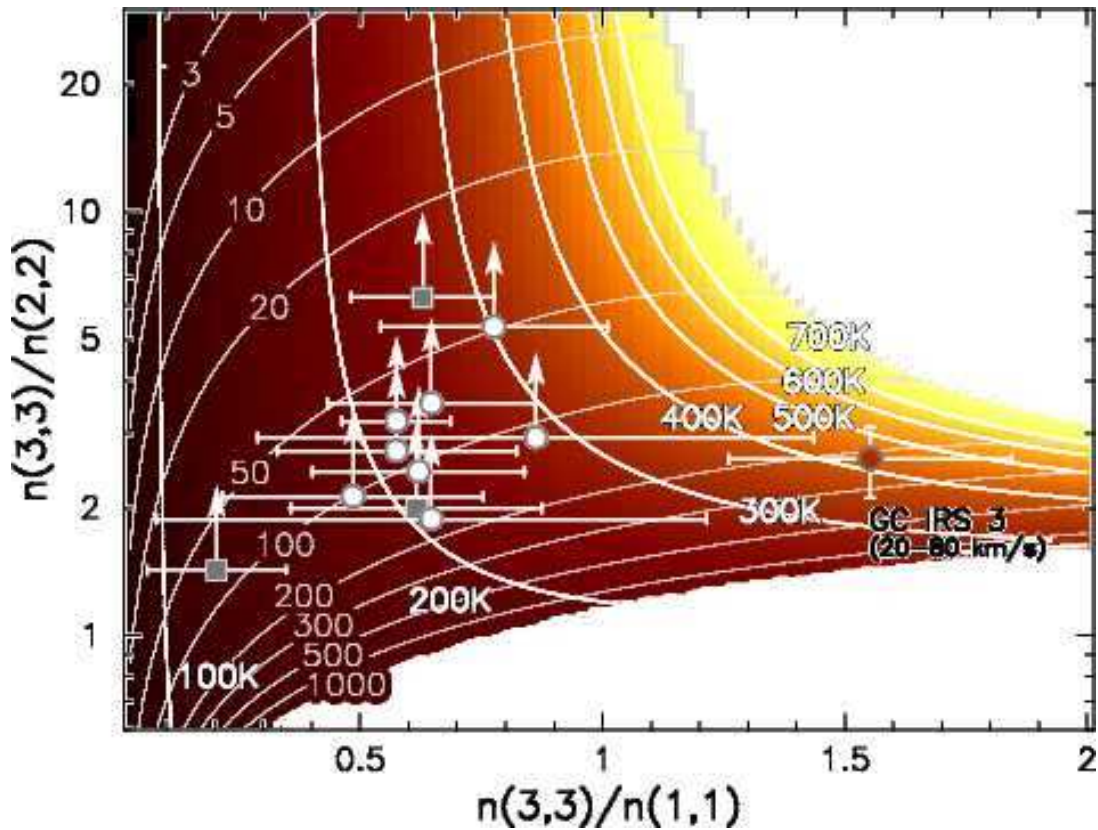}
\caption{Temperature and density plotted in filled circles as
functions of population ratios $n$(3,3)/$n$(1,1) and
$n$(3,3)/$n$(2,2) obtained by inverting the diagram of
\citet{oka04}. Temperature is indicated by thick white lines and
density by thin lines. The estimated uncertainties in
$n$(3,3)/$n$(1,1) are shown in error bars. Velocity resolved data of
GCS~3-2 from \citet{oka05} is shown in filled squares for
comparison.
\label{f4}}
\end{center}
\end{figure*}

\subsection{The Case of High $\zeta$ in the CMZ}

It is usually assumed that the density of high energy ($E > 100$~MeV)
cosmic rays is approximately uniform over the Galaxy because of their
long mean free paths. The uniform distribution on the sky of diffuse
$\gamma$ ray emission appears to lend support to this
\citep[e.g.][]{die01}. Recent detailed calculations by \citet{bue05}
typically give a 20~\% variation depending on location and
time. However, the situation is very different for low-energy cosmic
rays, which are the most important for ionization of the gas; their
energy densities can vary greatly depending on proximity to local
cosmic sources \citep{kul71,spi75,ces75} because of their short mean
free paths. For example, a proton of 2~MeV, which effectively ionizes
hydrogen, travels only 3~pc before losing its energy in a medium with
$n_H$ = 100~cm$^{-3}$ \citep{cra78}.  The ionization rate by 
low-energy cosmic rays could be proportional to the number density of
the accelerating sources \citep[SNRs; e.g.][]{koy95} and therefore
have strong local variations. If the energy density of cosmic-rays
follows the surface density of OB stars, the cosmic ray ionization
rate would be an order of magnitude higher in the Galactic center
than in the solar neighborhood \citep{wol03}.  \citet{yus02} and
\citet{yus07} called for an even higher local enhancement of
low-energy cosmic rays in the Galactic center ($\zeta =
2~\times~10^{-14}$ s$^{-1}$ to $5~\times~10^{-13}$s$^{-1}$!) to
account for the iron fluorescence line at 6.4~keV. Such values are
much larger than those derived in the previous section using
H$_3^+$. X-rays and far ultraviolet radiation from stars may also
increase $\zeta$. A factor of 10 or higher increase of $\zeta$ due to
X-ray emission has even been reported in the circumstellar disk of a
low mass star \citep{dot04}.  The enhancements could be large for the
CMZ, where there are abundant intense emitters of X-rays and EUV
radiation.

For the above high values of $\zeta$, $L$ would be quite short
if Eq.~(2) is applied. However, there are two arguments against
adopting such high values of $\zeta$. First, as argued earlier,
the observed high velocity dispersion of H$_3^+$ indicates that
the gas is quite extensive rather than confined in many small
clumps. This indicates that there are non-local sources of
ionization covering an extensive region.

Second, the results of Eqs.~(1) and (2) are based on a simple linear
formalism, which is a good approximation for low $\zeta$. For much
higher $\zeta$ this should be replaced by higher order non-linear
formalism in which $N$(H$_3^+$) no longer increases proportionally to
$\zeta$, but tends to saturate. One source of such nonlinearity,
revealed by \citet{lis06,lis07}, is the dissociative recombination
of H$_2^+$, H$_2^+$ + e$^-$ $\rightarrow$ H + H, which competes with
the H$_3^+$ production reaction H$_2^+$ + H$_2$ $\rightarrow$ H$_3^+$
+ H. The former reaction, in addition to the dissociative
recombination of H$_3^+$, reduces the H$_3^+$ concentration twice. It
also increases $n$(H) and decreases $f$, hence the
non-linearity. In an example given in the lower figure of Fig.~3 of
\citet{lis06}, a ten times increase of $\zeta$ from 10$^{-16}$
s$^{-1}$ to 10$^{-15}$ s$^{-1}$ leads to only a fourfold increase of
H$_3^+$ (the saturation is more severe for the upper figure). The
importance of saturation depends sensitively on the value of
$f$, the fractional abundance of H$_2$. Another source of
non-linear behavior is the production of H through dissociation of
H$_2$ by cosmic rays, X-rays, and EUV radiation. The rates of these
processes are comparable to the rate of ionization
\citep{gla73,gla74}. Since the reproduction of H$_2$ on dust grains is
slow, H accumulates in the gas and the charge exchange reaction,
H$_2^+$ + H $\rightarrow$ H$_2$ + H$^+$, which has a high rate
constant of 6.4~$\times$~10$^{-10}$~cm$^3$~s$^{-1}$, will compete with
the H$_3^+$ producing reaction introducing additional
non-linearity. The increase of $n$(H) and resulting reduction of
$f$ affect the gas in several other ways, all of which make the
increase of $\zeta$ ineffective in increasing
$\mathrm{H}_{3}^{+}$. While more detailed treatments of these
nonlinearities have yet to be worked out, it seems unlikely to us that
$\zeta$ can be larger than $\sim$ 3 $\times$ 10$^{-15}$ s$^{-1}$ over
extensive portions of the CMZ. Small volumes with higher $\zeta$ may
well exist in the CMZ, but they would not necessarily produce a large
fraction of the observed H$_{3}^{+}$.



\subsection{The $+$50~km~s$^{-1}$ absorption toward GC~IRS~3}

The sight line toward GC~IRS~3 is special in that at
$+$50~km~s$^{-1}$ LSR it has a detectable population of H$_3^+$ in
the unstable (2,2)~level in addition to the large populations in the
(1,1) and (3,3) levels. Also this is the only case for which the
population in the (3,3) metastable level is definitely higher than
in the (1,1) ground level (see $N$(3,3)/$N$(1,1) in Table 4). The
observable population in the (2,2) level indicates a high density
and the larger population in the (3,3) level than in the (1,1) level
indicates high temperature environment. Our analysis yields
$n$~=~175~--~300 cm$^{-3}$ and $T$~=~350~--~500~K. As seen from 
Fig.~4, this environment is clearly distinct from those elsewhere. 
We speculate below on two possible locations for this gas: (1) the
Circumnuclear Disk (CND) and, (2) the ``50~km~s$^{-1}$ cloud''
behind the GC.

\subsubsection{Circumnuclear disk}

GC~IRS~3 is only 5\arcsec~(0.2~pc) from Sgr~A$^\ast$, which is well within
the projected span of the Central Cluster \citep[the maximum
transverse separation of bright stars in the cluster is $\sim$0.65~pc;
][]{vie05}. The central 3~pc of the GC contains hot, high density gas
which shows intricate mini-spiral structures \citep{eke83,lo83} termed
the Western Arc, the Northern Arm, the Eastern Arm, and the Bar
\citep{gue87}. The extension of the Western Arc has been associated 
with a nearly completely
circular, Keplerian circumnuclear disk by analysis of HCN radio emission
\citep{gue87,gen87,jac93,chr05} and by other methods; recent direct
observations of the disk via its far infrared dust emission by
\citet{lat99} are particularly noteworthy.
There also have been attempts to interpret other three mini-spirals 
as CNDs which are not necessarily elliptical \citep[][e.
g.]{lis85,qui85,lac91,jac93,lis03}. GC~IRS~3 is located in the
cavity between the Northern Arm and the Bar, a region of low gas
density. H. S. Liszt (private communication) has suggested that the
\ion{H}{1} absorption at $+$50~km~s$^{-1}$ in Fig.~5 of
\citet{lis85} may be due to the same cloud producing the H$_3^+$
absorption in the sightline toward GC~IRS~3. Not only does the velocity
match, but also the high velocity gradient of 100~km~s$^{-1}$
arcmin$^{-1}$ found by \citet{lis85} is consistent with the width of the
H$_3^+$ 50~km~s$^{-1}$ absorption feature.


However, there are two possible problems with this interpretation.
First, it calls for an uncomfortably high cosmic ionization rate in
order to account for the observed H$_3^+$ column density in the
short path length within the CND. Assuming a path length on the
order of 0.5 pc, the observed ($\zeta$$L$)$_{\rm min}$ of 7.5
$\times$ 10$^4$~cm~s$^{-1}$ in Table~\ref{tb4} gives
$\zeta$~$\sim$~2~$\times$~10$^{-14}$~s$^{-1}$, an extremely
high value.  Such a value is possible \citep{yus07} in view
of the high density of stars and radiation field in the region, but
in such conditions the higher order chemistry \citep{lis06,lis07}
discussed in Section 5.3. surely plays a role. Although the
detection of H$_3^+$ in the unstable (2,2) level indicates higher
density compared with other sight lines, the population of the (3,3)
level is still much higher and the distribution is highly
non-thermal. The density is definitely much lower than that reported
for the CND (10$^4$~--~10$^6$~cm$^{-3}$). This could be due to the
fact that GC~IRS~3 is in the northern cavity and is only behind the
edge of the CND rather than its densest regions.

The second problem is the absence of the $R$(2,2)$^l$ line in the
spectrum of GC~IRS~1W. It is remarkable that the $R$(2,2)$^l$ line
is detected toward GC~IRS~3, but not toward GC~IRS~1W which is just
8\farcs4 apart, or 0.31~pc for a common distance of 7.6~kpc
\citep{eis05,nis06}. If both stars belong to the Central Cluster
\citep{bec78,vie05}, it is not likely that they are much more than
0.5~pc apart. Although the source might well be located in the
foreground, GC~IRS~1W overlaps with the high density part of the
Northern Arm, while GC~IRS~3 is located in the cavity. It is difficult
To understand why only GC~IRS~3 shows $R$(2,2)$^l$ absorption, if the
absorption occurs within the CND.


\subsubsection{The``50~km~s$^{-1}$ cloud''}

The observed velocity of $+$50~km~s$^{-1}$ toward GC~IRS3 immediately 
suggests that the
absorption might arise in the well known ``50~km~s$^{-1}$ cloud'',
one of the giant molecular clouds ($> 10^6 M_\odot$) located near
the Galactic center. The ``50~km~s$^{-1}$ cloud'' is sometimes
referred to as M$-$0.02$-$0.07, or was called the ``40~km~s$^{-1}$
cloud'' by \citet{oor77}, and has been discussed in detail by
\citet{bro84}.
 Many atomic and molecular studies have established that it is located
\emph{behind} the Galactic nucleus, in between the non-thermal radio
source Sgr~A East (SNR) and the \ion{H}{2} region Sgr~A West, which
enshrouds the central star cluster
\citep{whi74,gue83,pau96}. \citet{yus96} argued that the several
masers they discovered in the southeast of Sgr~A$^\ast$ are excited in
the shock-heated interface between Sgr~A~East and M$-$0.02$-$0.07,
where the expanding supernova remnant hits the giant molecular
cloud. \citet{tsu06} also found clues for an interaction between
Sgr~A~East and the ``50~km~s$^{-1}$ cloud'' in the high velocity wings
of CS $J=1-0$, although it may not be the case that the
``50~km~s$^{-1}$ cloud'' is the sole cloud that falls in the radial
velocity $+$50~km~s$^{-1}$ in the line of sight toward the Galactic
center \citep{zyl92}.

\citet{geb89} found that GC~IRS~3 and IRS~7 show deep absorption lines
of the CO fundamental band at 4.7~$\mu$m at the radial velocity
$+$50~km~s$^{-1}$, while GC~IRS~1 and IRS~2 do not. They attributed
the $+$50~km~s$^{-1}$ absorption to the CND but also mentioned the
possibility that GC~IRS~3 is not a member of the central cluster, but
behind the $+$50~km~s$^{-1}$ cloud and thus not in the Central 
Cluster. The nature of GC~IRS~3 is
controversial \citep{pot08}; it is among the brightest sources
in the Central Cluster region only less lumnous than GC~IRS~1 
and GC~IRS~10, and much redder than them in the mid-infrared 
even though it is (apparently) in the
cavity. Because of its location and featureless infrared continuum
\citep[e.g. ][]{kra95}, GC~IRS~3 has widely been believed to be a hot
massive star heavily obscured by its own dust (similar to the 
bright Quintuplet sources), in addition to 
foreground dust. However, new interferometric observation that 
resolve the circumstellar dust shell have been interpreted by 
\citet{pot08} as evidence that the central object is  a late-type 
carbon star, the photospheric spectral features of which are completely 
obscured. This interpretation is consistent with the apparent lack 
of ionization of the circumstellar shell, which one might expect 
if the object lies within the Central Cluster, and indeed the lack of 
ionized gas near GC~IRS~3 \citep{roc85}. Although somewhat unlikely
due to its proximity on the plane of the sky to the Central Cluster,
the placement of GC~IRS~3 well behind the cluster and the 
``$+$50~km~s$^{-1}$ cloud'' is otherwise a straightforward assignment 
and avoids the difficulties associated with the CND hypothesis 
mentioned in the previous section.



\subsection{Infrared pumping}

\begin{figure}
\begin{center}
\includegraphics[angle=0,width=.4\textwidth]{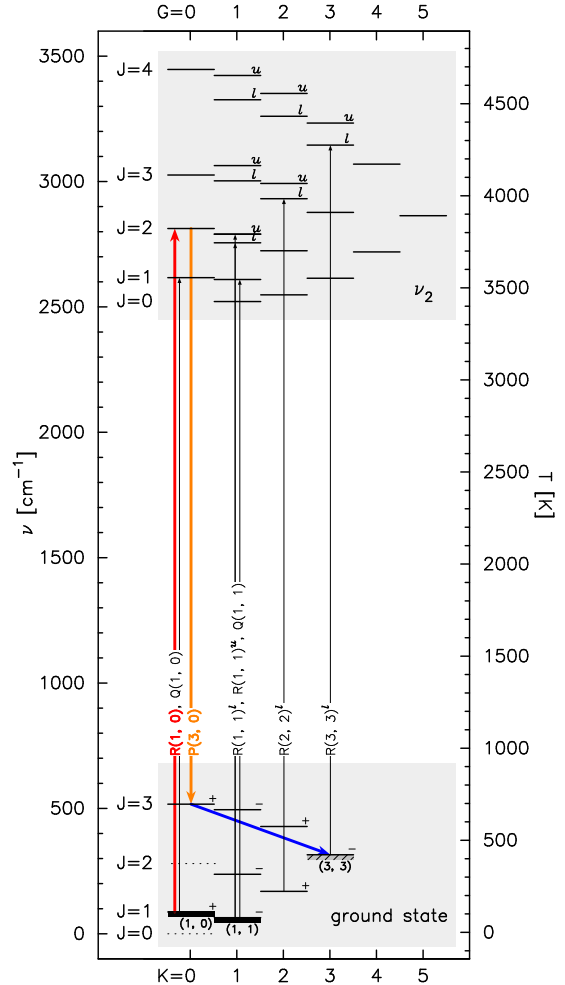}
\caption{Schematic of the infrared pumping mechanism proposed by J.
H. Black. H$_3^+$ in the lowest ortho-level (1,0) is excited to the
(2,0) level in the first excited vibrational state via the $R$(1,0)
transition (red line) by ambient photons. The excited H$_3^+$
spontaneously decays to the (3,0) ground level through the P(3,0)
transition (orange line) and then to the metastable (3,3) level
(shown in blue). See text for more details. Other transitions used
in this paper are also shown. \label{f5}}
\end{center}
\end{figure}

During the 2005 Royal Society Discussion Meeting on H$_3^+$
\citep{oka06a}, J. H. Black proposed an infrared pumping mechanism
which can produce a non-thermal rotational level distribution of
H$_3^+$ if the infrared flux is sufficiently high. The mechanism is
illustrated in Fig.~\ref{f5}. Infrared radiation pumps H$_3^+$ from
the fully populated lowest ortho-level, (1,0), to the (2,0) level in
the first excited vibrational state via the $R$(1,0) transition. The
molecule decays to the (3,0) level in the ground state by spontaneous
emission with a lifetime of 25.3~ms. The molecule further decays to
the (3,3) metastable level through the forbidden rotational transition
(3,0) $\rightarrow$ (3,3) emitting a far-infrared photon at 49.62
$\mu$m in 3.75 hours \citep{pan86,nea96}.  Thus, H$_3^+$ molecules are
pumped from the (1,0) level to the metastable (3,3) level.

Since the spontaneous emissions are all very fast, the rate
determining process for Black's mechanism is the rate of infrared
pumping of the (1,0) level of H$_3^+$. We rewrite the excitation
rate given in Eq.~(15) of \citet{dis82} as
\[
 W_{R(1,0)} = \lambda^2A\phi_\nu/8\pi = 1.51 \times
10^{-7} \phi_\nu {\rm s}^{-1},
\]

\noindent where $\phi_\nu$ is the photon flux of the ambient radiation
at the wavelength of the transition in the units of photon cm$^{-2}$
s$^{-1}$ Hz$^{-1}$, and the values of $\lambda$ = 3.6685~$\mu$m, the
Einstein coefficient $A$ = 98.5~s$^{-1}$, and the dilution factor of
0.286 are used. The dilution factor takes into account the spontaneous
emission (2,0) $\rightarrow$ (1,0) back to the ground level (1,1) with
$A$ = 35.5 s$^{-1}$, resulting in a reduction of the pumping
efficiency. In order for pumping to be significant, the value of $
W_{R(1,0)}$ has to be comparable to the collision rate,
\[
R = n\sigma v = n k_L \sim 2 \times 10^{-7}
{\rm s}^{-1},
\]
\noindent where we used a number density of $n$ = 100 cm$^{-3}$ and
the temperature independent Langevin rate constant,
2~$\times$~10$^{-9}$~cm$^3$~s$^{-1}$ for the H$_3^+$ + H$_2$
collision.

The energy density of the interstellar radiation field in the solar
neighborhood at 4~$\mu$m has been estimated by \citet{por06} to be
$\lambda u_\lambda$ = 0.055 $\mu$m~eV~cm$^{-3}$~$\mu$m$^{-1}$ which
translates to $\phi_\nu$~=~6~$\times$10$^{-5}$
photons~cm$^{-2}$~s$^{-1}$~Hz$^{-1}$. This is consistent with the
value reported by \citet{mat83} within a factor of 1.5. The
corresponding pumping rate is $W_{R(1,0)}$~=~1~$\times$~10$^{-11}$
s$^{-1}$, about 2$\times$10$^{4}$ times lower than the collision rate.
\citet{por06} have estimated an average energy density about 20 times 
higher in the GC. \citet{lau02} have reported an energy density in 
the central
120~pc (their Fig.~10) which is another factor of 10 higher. The
corresponding pumping rate is still 100 times lower than the collision
rate and the infrared pumping is perhaps not very effective for
majority of the gas reported in this paper.

In the central parsec of the Galaxy, however, the stellar density is
$\sim$ $10^6$~$M_\odot$~pc$^{-3}$ \citep{sch07}, including a large
number of luminous red giants and at least 80 hot and massive stars
\citep{pau06}. In this environment infrared pumping likely would
overwhelm the collisional relaxation of any H$_3^+$ that is present.
The effect, however, must be local to the very central region of the
CMZ and probably does not seriously alter the population of the
(3,3) level beyond a few tens of parsecs from the center, i.e., in
the bulk of the CMZ. However, it may well be the best explanation
for the strong $R$(3,3)$^l$ absorption line toward GC~IRS~3 at
50~km~s$^{-1}$ discussed in the previous subsection.

\section{Summary Points}

(1) Seven infrared spectral lines of $\mathrm{H}_{3}^{+}$, $Q$(1,0),
$Q$(1,1), $R$(1,1)$^l$, $R$(1,0), $R$(1,1)$^u$, $R$(2,2)$^l$, and
$R$(3,3)$^l$, have been observed toward eight bright and hot stars
distributed from the Quintuplet Cluster $\sim$ 30 pc east of the
Galactic Center, to the Central Cluster close to Sgr~A$^\ast$, using
the IRCS spectrograph of the Subaru Telescope.

(2) All sight lines showed strong $R$(3,3)$^l$ absorptions with high
velocity dispersions, indicating a high population of the (3,3)
metastable rotational level. Those absorptions are ascribed to
$\mathrm{H}_{3}^{+}$ in the CMZ and indicate a large surface filling
factor.

(3) The five absorption lines starting from the lowest levels, (1,1)
and (1,0), are composed of sharp spectral lines due to
$\mathrm{H}_{3}^{+}$ in the foreground spiral arms and a broad
trough due to $\mathrm{H}_{3}^{+}$ in the CMZ. These two 
absorption components have been separated by assuming that the 
profile of the trough
is proportional to the $R$(3,3)$^l$ absorption profile.

(4) The observed high $\mathrm{H}_{3}^{+}$ column densities indicate
lower limits of $\zeta$$L$, ($\zeta$$L$)$_{\rm min}$ = (1.3 -- 4.5)
$\times$ 10$^5$ cm s$^{-1}$, which are 1000 and 10 times higher than
for dense and diffuse clouds in the Galactic disk, respectively,
indicating high values of $\zeta$ and $L$ in the CMZ.

(5) Various arguments suggest $\zeta$ is on the order of 10$^{-15}$
s$^{-1}$ and $L$ on the order of 50 pc in the CMZ, indicating a high
ionization rate in the region and a high volume filling factor of
the gas.

(6) Analysis based on the thermalization model calculation
\citep{oka04} has given $T$ = 200 -- 300 K and
$n$~$\leq$~50~--~200~cm$^{-3}$ (Fig.~\ref{f4}). A lower limit 
on the order of 50 cm$^{-3}$ is set for $n$ from the observed visual
extinction, $A_V$, and the existence of detectable
$\mathrm{H}_{3}^{+}$.

(7) Taken as a whole the observations reveal a previously
unidentified gaseous environment in the CMZ having a high volume
filling factor. This is radically different than the previous picture 
of the gas in the CMZ, and suggests in particular that the filling 
factor of ultra-hot X-ray emitting plasma may be nowhere close to
unity.

8)~ The sight line toward GC~IRS~3, close to Sgr~A~$^\ast$, is
unique in having a detectable population in the H$_3^+$ (2,2)
unstable level near $+$50~km~s$^{-1}$, indicating a higher density
for the gas producing this absorption. We speculate that the 
cloud responsible for
this absorption feature is either a compact cloud near the
circumnuclear disk or the well known ``50~km~s$^{-1}$ cloud''.  For
the former a high value of $\zeta = 2 \times 10^{-14}$ s$^{-1}$ is
needed, while for the latter GC~IRS~3 is located behind the  
``50~km~s$^{-1}$ cloud'' and thus outside of the Central Cluster.

(9) An infrared pumping mechanism to populate the (3,3) metastable
level proposed by J. H. Black has been examined. The infrared
radiation in the extended CMZ is not intense enough to significantly
pump this level, and the high temperatures detected there are likely
thermal.  However in the CND pumping probably dominates and 
temperatures determined for gas in that region based on the 
relative populations in this level and the $J$~=~1 levels would 
not be kinetic temperatures.


\vspace{0.15in}

We thank all the staff of the Subaru Telescope and NAOJ for their
invaluable assistance in obtaining these data and for their continuous
support during IRCS and Subaru AO construction. We are grateful to the
staff of UKIRT, which is operated by the Joint Astronomy Centre on
behalf of the U. K. Science and Facilities Research Council, for
assistance in obtaining the corroborating spectra of GC~IRS~3. Special
thanks go to John H. Black, Katsuji Koyama, Harvey S. Liszt, Kazuo
Makishima, Tomoharu Oka, Takeshi Tsuru, and Farhad Yusef-Zadeh for
their enlightening discussions and to Joseph Lazio and H. S. Liszt for
critical reading of this paper.  This research has made use of the
SIMBAD database, operated at CDS, Strasbourg, France. T. R. G.'s
research is supported by the Gemini Observatory, which is operated by
the Association of Universities for Research in Astronomy, Inc., on
behalf of the international Gemini partnership of Argentina,
Australia, Brazil, Canada, Chile, the United Kingdom and the United
States of America. M. G. was supported by a Japan Society for the
Promotion of Science fellowship. B. J. M. and N. I. have been
supported by NSF Grant PHY 05-55486. T.O. acknowledges NSF Grant PHY
03-54200. We wish to express our appreciation for the hospitality of
Hawaiian people and the use of their sacred mountain.



\end{document}